# Determining Quasi-Equilibrium Electron and Hole Distributions of Plasmonic Photocatalysts using Photomodulated X-ray Absorption Spectroscopy

Levi D. Palmer[1], Wonseok Lee[1], Chung Li Dong[2], Ru-Shi Liu[3], Nianqiang Wu[4], Scott K. Cushing[1,*]

[1]Division of Chemistry and Chemical Engineering, California Institute of Technology, Pasadena, CA 91125, USA.

[2]Department of Physics, Tamkang University, New Taipei City 251301, Taiwan

[3]Department of Chemistry, National Taiwan University and Advanced Research Center for Green Materials Science and Technology, Taipei 10617, Taiwan

[4]Department of Chemical Engineering, University of Massachusetts Amherst, Amherst, MA 01003–9303, United States

*Corresponding author. Email: scushing@caltech.edu

KEYWORDS: photomodulated X-ray spectroscopy, ab initio theory, plasmonic nanoparticles, hot carriers

**ABSTRACT:**

Most photocatalytic and photovoltaic devices operate under broadband, constant illumination. Electron and hole dynamics in these devices, however, are usually measured using ultrafast pulsed lasers in a narrow wavelength range. In this work, we prove that steady-state, photomodulated X-ray spectra from a non-time-resolved synchrotron beamline can be used to estimate electron and hole distributions. A set of plasmonic metal core-shell nanoparticles is designed to systematically isolate photothermal, hot electron, and thermalized electron-hole pairs in a $TiO_2$ shell. Steady-state changes in the Ti $L_{2,3}$ edge are measured with and without continuous-wave illumination of the nanoparticle's localized surface plasmon resonance. *Ab initio* excited-state X-ray theory developed for transient X-ray measurements is then applied to model the experimental spectra in an attempt to extract the resultant steady-state carrier distributions. The results suggest that, within error, the quasi-equilibrium carrier distribution can be determined even from relatively noisy data with mixed excited-state phenomena.

**TOC Figure:**

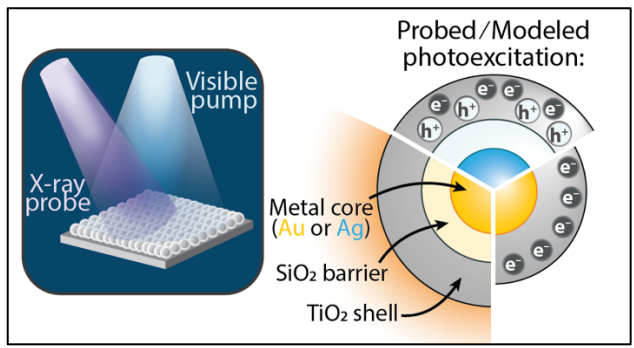

A balance between carrier photoexcitation, thermalization, and recombination rates determines the quasi-equilibrium carrier distribution that controls photocatalytic and photovoltaic device efficiencies (**Figure 1**).[1–3] A quasi-equilibrium state occurs during the thermodynamic balance of the system's photoexcitation and relaxation. Photoexcited carriers are generally assumed to be fully thermalized to the band edges at the device's working conditions.[4] However, slowed hot carrier cooling through phonon bottlenecks,[5] surface state trapping,[6–9] or dielectric carrier Coulomb screening[10] can generate a non-thermal carrier quasi-equilibrium. In nanoscale junctions, photoexcited carriers can transfer between active layers or to surface catalysts on timescales shorter than carrier thermalization.[11,12] Transferring the quasi-equilibrium hot carrier population into surface reactants then modifies a semiconductor's photochemical redox potential, tailoring resultant reaction products.[13] For example, plasmonic metal-semiconductor junctions have been used to increase semiconductors' photocatalytic product selectivity[14–16] and solar power conversion efficiency.[17,18]

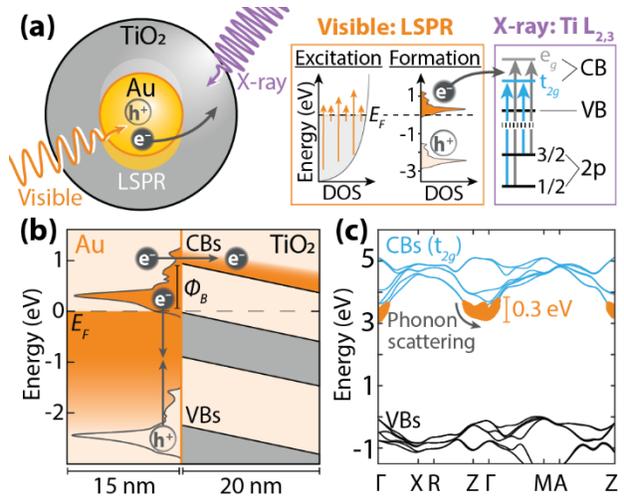

**Figure 1. Photoexcited properties of plasmonic core-shell nanoparticles.** (a) Continuous photoexcitation of a metal nanoparticle's localized surface plasmon resonance (LSPR) with visible light results in dynamic carrier excitation.[ref. 2] Hot carrier formation then occurs following electron-electron and electron-phonon scattering.[ref. 3] The hot carriers transfer into $TiO_2$ and can be probed with the Ti $L_{2,3}$ edge. (b) The hot carriers transfer over the Au@$TiO_2$ Schottky barrier ($\phi_B$) and fill the $TiO_2$ conduction bands (CBs) with an excess energy of 0.3 eV and subsequently (c) thermalize in the CBs through phonon scattering.

Measuring the quasi-equilibrium carrier distribution is therefore critical. However, few methods to date can characterize the equilibrium photoexcited carrier population with the same detail as ultrafast pump-probe methods like two-dimensional, terahertz, or photoemission spectroscopies.[19–22] While ultrafast spectroscopy is the conventional method for measuring carrier thermalization and recombination, ultrafast measurements sum over different relaxation pathways, often use a high peak power that exceeds the solar flux, and rely on laser sources that are more narrowband than the solar spectrum. When effects such as Fermi level pinning, defects, and surface states are present, it can be difficult to reconstruct steady-state carrier distributions using ultrafast measurements alone.

X-ray spectroscopy is one potential method for resolving carrier distributions and dynamics. Transient X-ray spectroscopy is now routinely used to measure element-specific electron and hole energies in multi-element catalysts.[23–26] The same capabilities should also be true for steady-state, photomodulated X-ray spectroscopy.[27] However, measuring and interpreting photomodulated X-ray spectra are challenging tasks because the decreased photo-induced carrier concentration and slower repetition rate make the signal-to-noise ratio significantly lower. Therefore, accurate excited-state X-ray theory is needed, even more so than for ultrafast X-ray spectroscopy, to interpret the small photomodulated spectral intensity within the experimental noise.

Previous investigations of plasmonic Au@$TiO_2$ nanoparticles and their photomodulated X-ray spectra suggest that quasi-equilibrium hot electrons populations exist in $TiO_2$.[7,28,29] In our prior work, the photoexcited X-ray spectra were not modeled fully *ab initio* but were rather calculated using a semi-quantitative model of phase-space filling and lifetime effects.[29] Thermal and photoexcited hole effects were not included.

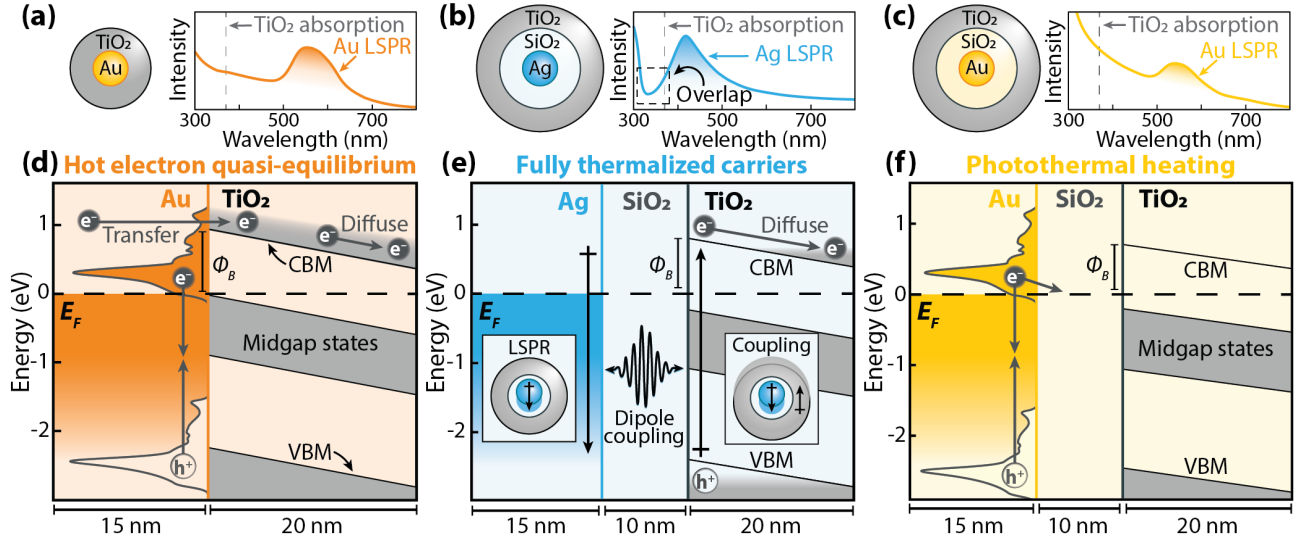

**Figure 2. Plasmonic core-shell nanoparticle heterostructure characterization.** (a-c) UV–visible absorption spectra for (a) Au@TiO$_2$, (b) Ag@SiO$_2$@TiO$_2$, and (c) Au@SiO$_2$@TiO$_2$ core-shell nanoparticles. The localized surface plasmon resonance (LSPR) and UV bandgap for amorphous TiO$_2$ (grey dashed line at 3.34 eV, 370 nm) are marked [refs. 43,44]. (d-f) Schematic representations of each nanoparticle's band alignment and hot carrier distribution (d) Hot electrons up to ~0.3 eV above the conduction band minimum (CBM), relative to the Schottky barrier ($\phi_B$), have sufficient energy to transfer into TiO$_2$ directly.[ref. 3] (e) The Schottky barrier and SiO$_2$ layer prevent hot electron transfer. Instead, the localized electromagnetic field from the plasmon couples with a TiO$_2$ electron-hole excitation, and carriers are created at both the CBM and valence band maximum (VBM). (f) The SiO$_2$ layer prevents electron transfer into TiO$_2$.

In this work, we test whether photomodulated, steady-state X-ray spectroscopy can be used to quantify quasi-equilibrium carrier distributions. X-ray absorption at the Ti L$_{2,3}$ edge is measured for each nanoparticle with modulated photoexcitation. An adiabatic approximation to the Bethe-Salpeter equation (BSE) is then used to predict the change in the X-ray spectrum for each possible photoexcited state. We show that, even in the case of relatively noisy spectra, quasi-equilibrium hot carrier distributions can be differentiated from photothermal effects. The hypothesized lattice temperatures and carrier distributions are tested using a mean squared error (MSE) analysis. Separating electron versus hole effects from a photothermal background was found to be more difficult because of the hole's smaller perturbation to the Ti L$_{2,3}$ edge. Due to experimental noise, electron versus hole populations were not separable with statistical significance, but the calculations do demonstrate that electrons and holes have distinct spectral effects and could be differentiated with improved X-ray measurements. These findings suggest that photomodulated X-ray spectroscopy at non-time-resolved beamlines can be used to separate electron, hole, and thermal effects.

## RESULTS AND DISCUSSION

A nanoparticle's localized surface plasmon resonance (LSPR) can be used to transfer energy to a semiconductor through multiple mechanisms. Here, a SiO$_2$ layer is used to systematically control three such mechanisms between a Au or Ag nanoparticle core and a TiO$_2$ shell. For Au@TiO$_2$ nanoparticles, plasmonic hot electrons in Au can overcome the interfacial Schottky junction to inject into TiO$_2$ (**Figure 2a,d**).[3,30,31] Ag@SiO$_2$@TiO$_2$ nanoparticles use the plasmon's dipole moment to increase the light absorption rate in the tail of a semiconductor's absorption edge, creating electron-hole pairs (**Figure 2b,e**).[20,28] In Au@SiO$_2$@TiO$_2$ nanoparticles, a SiO$_2$ layer prevents carrier transfer from Au into TiO$_2$, so the TiO$_2$ shell only experiences heating from the Au core to provide a control experiment (**Figure 2c,f**).[28] Past work has verified that these core-shell nanoparticles isolate these excited-state effects.[28]

This paper theoretically interprets previously measured X-ray spectra of the nanoparticles synthesized and characterized in Ref. 28. In these previous measurements, the nanoparticles were reported to have a 15 nm radius Au or Ag core, a 10 nm SiO$_2$ insulating layer (not present in Au@TiO$_2$), and a 10 – 20 nm amorphous TiO$_2$ outer shell. UV–visible absorption spectroscopy was used to measure the LSPR center wavelength at 420 nm for Ag and 560 nm for Au (**Figure 2a-c**).

The approximate interfacial band bending of each heterojunction is calculated using a 1D drift-diffusion model implemented in the Automat FOR Simulation of HETerostructures (AFORS-HET) (**Figure 2d-f**).[32] This approach does not consider nanoscale near-field or photoexcited effects. The approximate Schottky barriers are 0.9 eV for Au@TiO$_2$, 0.8 eV for Ag@SiO$_2$@TiO$_2$, and 0.7 eV for Au@SiO$_2$@TiO$_2$. The metal-semiconductor junction produces band bending and built-in electric fields in the TiO$_2$ and SiO$_2$ layers. The average built-in field estimated for the semiconducting layers is ~$10^5$ V/cm. The SiO$_2$ insulator acts as a carrier tunnelling barrier between the metal and TiO$_2$, and the Schottky barrier in such cases refers to the energetic barrier for electron transfer at the SiO$_2$-TiO$_2$ interface. Considering the ~4.4 eV Au-SiO$_2$ Schottky barrier, which exceeds the maximum hot electron energy by 3.2 eV, Au hot carriers would need to tunnel through SiO$_2$ to reach TiO$_2$. Similar junctions with a 4.8 nm SiO$_2$ oxide were previously measured to have a <$10^{-10}$ A/cm$^2$ tunneling current at a $10^5$ V/cm applied bias.[33] Therefore, as experimentally observed, photoexcited electrons would not transfer to TiO$_2$ for the Au@SiO$_2$@TiO$_2$ system. See the Supporting Information for numerical input parameters and field calculation.

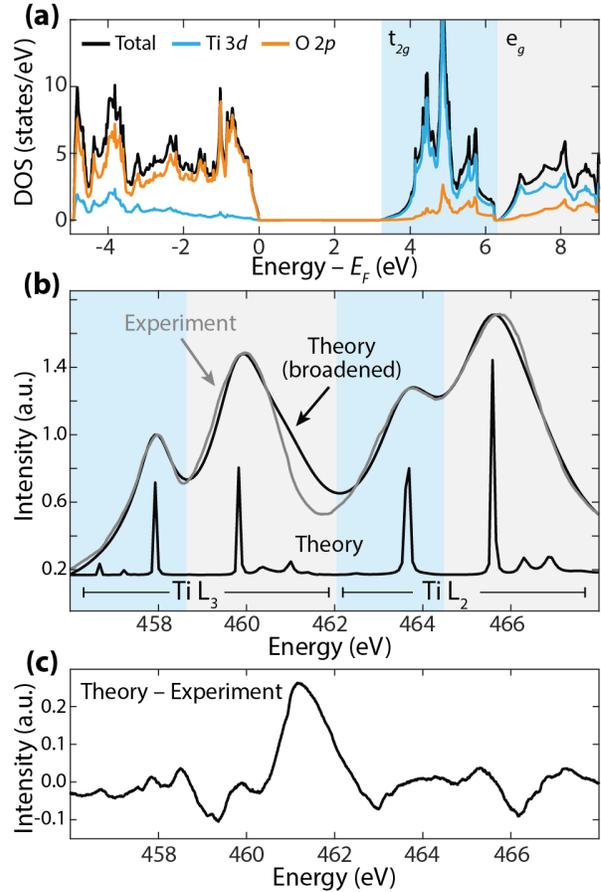

**Figure 3. Calculation of TiO$_2$ electronic structure and X-ray absorption.** (a) The DFT-calculated, projected density of states for anatase TiO$_2$. The Fermi level represents the valence band edge. (b) Calculated (black) and measured (grey) Ti L$_{2,3}$ X-ray spectra for TiO$_2$. The theory spectrum is broadened to match the experiment with the bottom spectrum being the unbroadened output. The blue and grey shading depict the Ti t$_{2g}$ and e$_g$ states, respectively. (c) The difference between theory and experiment.

The ground-state electronic structure and X-ray absorption of TiO$_2$ are first calculated as shown in **Figure 3**. The Ti L$_{2,3}$ X-ray absorption edge (456 – 468 eV) was measured, which corresponds to a core electron transition from Ti 2$p$ to Ti 3$d$ states (**Figure 1a, right**). The density functional theory (DFT) calculated projected density of states (PDOS) for anatase TiO$_2$ is given in **Figure 3a**. A 1 eV scissor shift is applied to the bandgap. In the PDOS, the O 2$p$ orbitals dominate the valence band and the Ti 3$d$ orbitals compose the conduction band. The crystal field characteristically splits the Ti 3$d$ conduction band into the $t_{2g}$ (blue shading) and $e_g$ (grey shading) orbitals in the electronic structure and ground-state X-ray absorption (**Figure 3a,b**).[34]

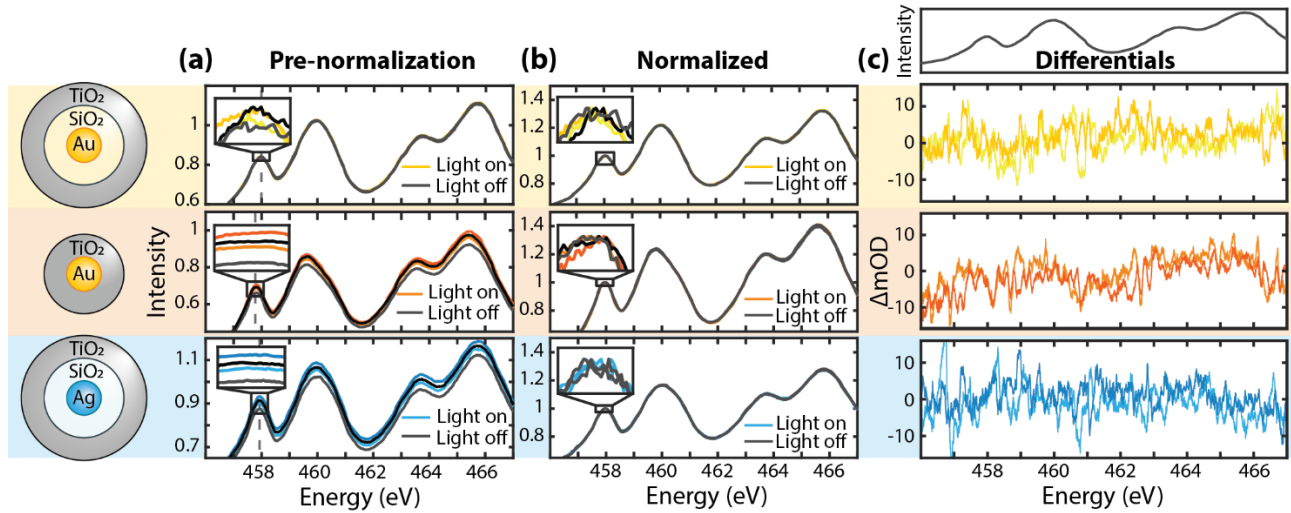

**Figure 4. X-ray absorption spectra of core-shell nanoparticles with and without photoexcitation.** (a) Raw, light on and light off Ti $L_{2,3}$ edge X-ray spectra of the amorphous $TiO_2$ outer shell. The spectral intensity increased throughout the data collection of all four spectra. The spectra are background-subtracted. Each inset magnifies the edge maxima intensity differences caused by charging. Each inset window size is 0.15 eV width but a variable amplitude. (b) Spectra from (a) normalized to the edge maximum near 458 eV (grey dashed line) to correct for charging that broadly increases the spectral amplitude. (c) Photoexcited differential spectra of the normalized data in (b) to highlight photomodulated energetic shifts in the $L_{2,3}$ edge. The lighter spectra are from the first light on/light off collection. The ground state experimental spectrum is shown above for reference.

**Figure 3b** compares the BSE simulated Ti $L_{2,3}$ edge to the measured ground-state $Au@SiO_2@TiO_2$ spectrum. An energy-dependent broadening method of the predicted spectrum was used to replicate the experimental core-hole lifetime broadening, which has a 3:2 broadening ratio ($L_2:L_3$) for the $TiO_2$ Ti $L_{2,3}$ edge (see Supporting Information).[35] According to the PDOS in **Figure 3a**, the photomodulated Ti $L_{2,3}$ edge predominantly probes photoexcited electrons over holes through the Ti $3d$ states; however, because of the screening and angular momentum coupling matrix elements in the X-ray transition Hamiltonian, holes will still perturb the core-to-valence transition excitons.[36,37]

This work approximates the nanoparticle's amorphous $TiO_2$ as purely anatase phase. Although 10 – 20 nm $TiO_2$ nanoparticles typically consist of a mixture of anatase and brookite, anatase is a slightly more stable phase, and this approximation reduces the otherwise insurmountable computational costs of excited-state X-ray BSE calculations for hundreds of atoms.[38–41] We find this to be a valid approximation due to the excellent match between the ground-state experiment and theory (**Figure 3b**). However, aspects of the amorphous phase electronic structure are not considered. First, there is a discrepancy between anatase and amorphous $TiO_2$ for the Ti $L_3$ $e_g$ states at 461 eV (**Figure 3c**).[42] The core-hole exciton effects calculated by the BSE are therefore not accurately modeled at these energies and are not considered during the MSE analysis. Further, defect-induced midgap states (depicted in **Figure 2d-f**) are not modeled. Midgap states show little effect on the ground-state $TiO_2$ spectra but may appear as a shoulder of the $L_2$ $t_{2g}$ peak at 463 eV (**Figure 3b,c**).

The previously-measured X-ray spectra were collected using a light-on, light-off sequence with a 1-minute collection time per spectrum. A continuous-wave lamp, filtered below the 3.34 eV amorphous $TiO_2$ bandgap, was used to photoexcite the nanoparticles' LSPR.[43,44] Surface charging from photoexcitation creates a baseline drift in photomodulated spectra for the total electron yield detection method.[45] To account for charging, each spectrum is normalized to the edge onset maximum near 458 eV (grey dashed line in **Figure 4**) after the baseline background subtraction. The charging normalization creates artifacts directly below and above the X-ray absorption edge, so only the 458 – 466 eV range is compared to theory herein (see average spectra for all samples overlaid in **Figure S5**).

The measured differential absorption, calculated as the log of spectra collected with the lamp on divided by those with the lamp off and averaged across two data sets, is used to identify photoexcited carrier and structural effects on the X-ray spectra (**Figure 4**). Because the total electron yield detection only probes the first ~4 nm of $TiO_2$, only carriers in surface trap states in the amorphous $TiO_2$ are probabilistically measured.[46] The spectra in **Figure 4** are relatively noisy due to the lower power of the excitation source and slower modulation time of the steady-state measurement. Therefore, we first test the accuracy of our *ab initio* approach by comparing to a previous ultrafast X-ray absorption spectrum of anatase $TiO_2$ (**Figure S1**).[27] The measured transient spectrum is analyzed using an adiabatic approximation to excited-state effects in the BSE. This approach has been verified previously for other transient X-ray datasets and is described in the **Methods** section.[26,36,37] The ultrafast time slice is after carrier thermalization (1 ps). The proposed *ab initio* method accurately reproduces the transient X-ray spectrum at all energies besides 458 – 460 eV. The discrepancy is likely due to the reported onset of carrier transfer to midgap states (see Supporting Information).

Given this verification, we proceed to analzye the photomodulated spectra. Within noise, three major differences are observed between each plasmonic excitation mechanism. First, all nanoparticle's spectra have different amplitudes just after the $L_3$ edge at 458 eV. Second, the Au@$TiO_2$ has decreased absorption centered at 462 eV. Lastly, the Ag@$SiO_2$@$TiO_2$ nanoparticles have decreased absorption at 465.5 eV.

The steady-state spectral signatures of photothermal effects are first tested using the Au@$SiO_2$@$TiO_2$ experimental control (**Figure 5**). Photothermal heating arises from the heat produced by carrier thermalization in the metal nanoparticle after LSPR photoexcitation and relaxation.[47] The photomodulated Au@$SiO_2$@$TiO_2$ nanoparticle's experimental spectra lack the surface charging artifact that results from photoexcited carriers in the other two nanoparticle systems (**Figure 4a**), confirming that photoexcited carriers are not excited in $TiO_2$.

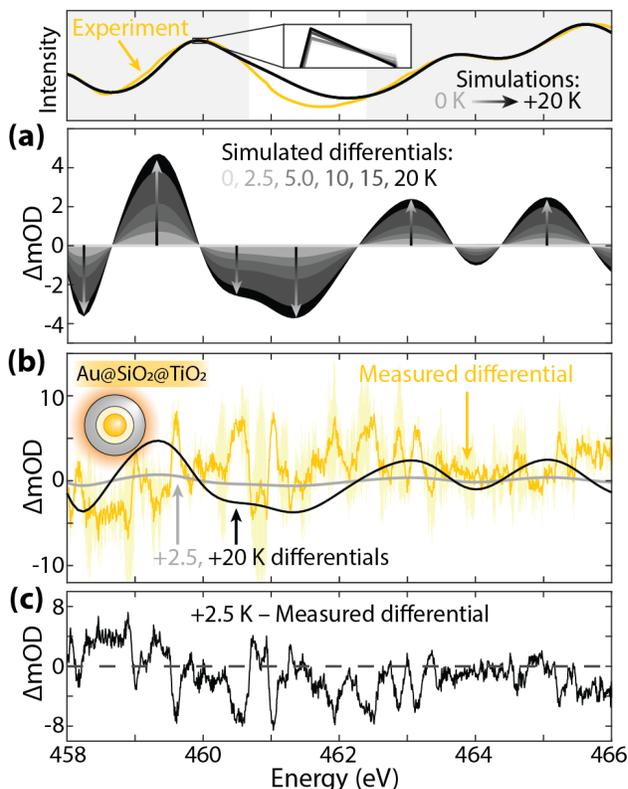

**Figure 5. Photothermal effects on the Ti $L_{2,3}$ edge in Au@$SiO_2$@$TiO_2$.** (Top) Raw spectra for reference. The spectral range analyzed by the MSE is shaded. (a) Simulated differential X-ray spectra for a 0 – 20 K anatase lattice expansion. The arrows highlight the differential amplitude change with an increasing lattice parameter. (b) The lattice-expanded differential spectra predicted for 2.5 and 20 K heating are compared to the Au@$SiO_2$@$TiO_2$ differential spectrum. The yellow shaded region depicts the experiment's standard deviation. (c) The measured differential subtracted from the optimized simulation differential at 2.5 K, selected using the MSE analysis.

Heating is modeled through DFT and BSE calculations by an expansion of the TiO$_2$ lattice. Calculations are performed for lattice expansions equivalent to 2.5, 5.0, 10, 15, and 20 K temperature increases above 300 K (**Figure 5a**).[48] The spectrum's peak positions linearly redshift with increasing lattice temperature, increasing and decreasing the differential intensity at pre- and post-edge regions, respectively. The spectrum mainly redshifts because of the Ti atoms' reduced crystal field. The simulated X-ray differential absorption for a 2.5 K lattice-expanded TiO$_2$ crystal is compared to the measured differential absorption in **Figure 5b**.

The Au@SiO$_2$@TiO$_2$ nanoparticle temperature after photoexcitation was predicted to be +2.5 K through a MSE fit of all simulations in **Figure S2**. Subtracting the 2.5 K heating differential from the experiment (**Figure 5c**) reflects that the general spectral changes are reproduced. However, the heating simulation's largest disagreement results from the anatase TiO$_2$ approximation at 460 – 462 eV. The predicted ~2.5 K rise in the TiO$_2$ layer is consistent with other studies reports of 7.7 K (theoretical)[49] and 2.6 ± 2.3 K (experimental)[50] heating of aqueous Au nanoparticles when using similar excitation densities. The thermal dissipation will of course differ in the vacuum environment for the experimental X-ray measurements.

Next, the Au@TiO$_2$ nanoparticle sample that has both photothermal and hot electron effects is examined. To simulate hot electron transfer in the Au@TiO$_2$ nanoparticles, electrons up to a specific energy above the TiO$_2$ CBM (0.0 eV for fully thermalized electrons, 0.1 eV, 0.3 eV, 0.45 eV, and 0.6 eV) are included in the BSE calculation (**Figure 6a**). We simulate spectra by approximating an average electron energy in the CBs and not a distribution. The added electrons change the core-hole screening and prevent X-ray transitions into the newly blocked states, leading to a complex differential absorption, as shown in **Figure 6b**. In **Figure 6b,** the intensity of the hot electrons' differential absorption is normalized to the total number of simulated hot electrons to better evaluate excited-state trends with increasing hot electron energy.

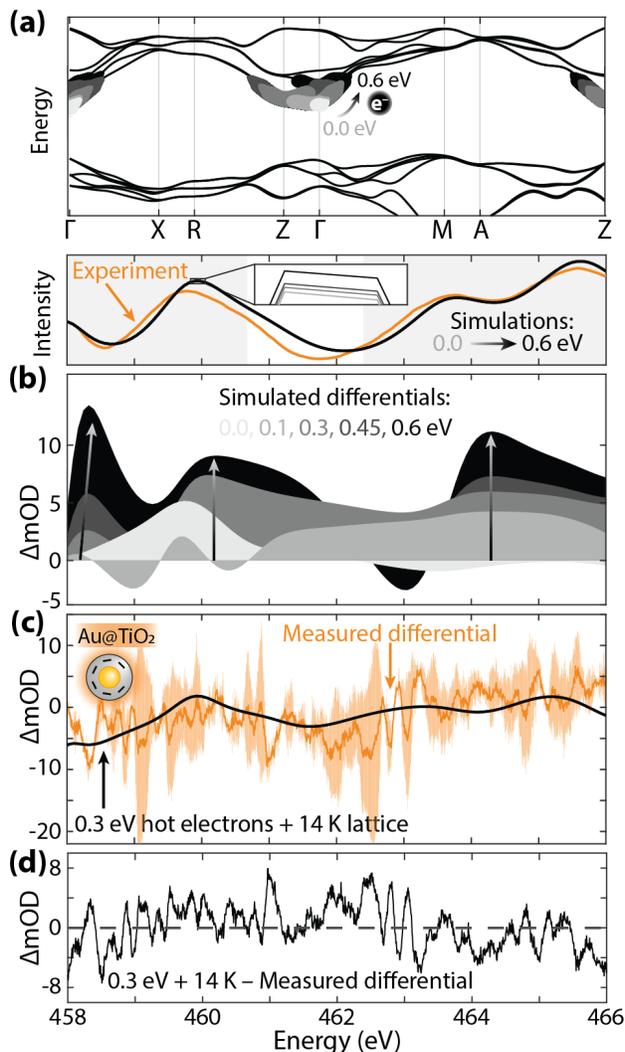

**Figure 6. Hot electron effects on the Ti L$_{2,3}$ edge in Au@TiO$_2$.** (a) Simulated hot electrons 0.0, 0.1, 0.3, 0.45, and 0.6 eV above the TiO$_2$ CBM (b, Top) Raw spectra for reference. The spectral range analyzed by the MSE is shaded. (b) Simulated differential X-ray spectra for each hot electron occupation in TiO$_2$. The spectral intensity is normalized to the number of electrons simulated in all but the thermalized (0.0 eV) case. (c) The optimized simulation with hot electrons 0.3 eV above the CBM with a 14 K lattice expansion is compared to the Au@TiO$_2$ measured differential absorption. The orange shaded region depicts the experiment's standard deviation. (d) The measured differential subtracted from the optimized simulation differential shown in (c).

Unlike photothermal heating, changes in TiO$_2$'s simulated differential absorption are not perfectly linear with increasing electron energy (**Figure 6b**). Instead, spectral intensity increases with the hot electron energy and there are differential peaks from changes in the screening of the core-valence exciton and X-ray transitions blocked by hot electrons. The differential peaks are mainly a result of hot electrons affecting the screening and angular momentum components of the core-valence exciton because the energies are well above the hot carriers at the bottom of the t$_{2g}$ bands (**Figure 6a**). The differential features blueshift as more hot electrons screen the exciton in the BSE. State-filling effects of hot electrons blocking X-ray transitions begin to appear at 463 eV when the simulated hot electrons fully occupy states above 0.45 eV.

The measured Au@TiO$_2$ nanoparticle differential absorption spectrum is compared to a simulated differential spectrum with 0.3 eV hot electrons and 14 K lattice expansion in **Figure 6c**. Compared to Au@SiO$_2$@TiO$_2$, the simulated X-ray spectrum with hot electrons has a new minimum around 462 eV, consistent with the measured spectral differences between samples with and without hot electrons (**Figure 4**). An MSE analysis was used to determine the most likely temperature and hot electron energy, based on the simulated spectra. Each modeled hot electron distribution is shown separately in **Figure S9**, and the differential X-ray spectra with both hot electrons and temperature simulated are in **Figure S11**. The lattice temperature is found to be hotter than the Au@SiO$_2$@TiO$_2$ nanoparticles, attributed to the larger extinction amplitude of the Au@TiO$_2$ particle's LSPR that would generate more carriers for recombination/heating (**Figure S3**). Further, the 0.3 eV hot electron quasi-equilibrium distribution is consistent with recent studies, as steady-state and ultrafast Raman measurements estimate that hot electrons exceeding 0.32 and 0.34 eV, respectively, transfer from Au nanoparticles to nearby molecules.[51,52] An approximate calculation comparing the electron excitation and relaxation rates in the amorphous TiO$_2$ is given in the Supporting Information, but the main conclusions of **Figure 6** are that photothermal and hot electron effects can be differentiated within a relatively noisy spectrum.

Further light intensity-dependent control experiments would be necessary to quantify the hot electron concentration. However, it can be approximated using the relative occupation of the state filling in the band structure. The state filling simulation for states up to 0.3 eV above the CBM best matches the experiment (**Figures 6c and S2b**). The integrated total DOS in this energy range is 150 states of the 10$^4$ total possible calculated conduction states. Using this number, and the measured DOS for nanocrystalline TiO$_2$ (8*10$^{18}$ cm$^{-3}$), one can approximate an electron concentration of ~10$^{16}$ cm$^{-3}$.[53]

Lastly, the Ag@SiO$_2$@TiO$_2$ nanoparticle system is examined and is expected to have thermalized electron and hole pairs (**Figure 7**). This is the most challenging example to model as electrons, holes, and photothermal effects are present within the noisy experimental spectrum. The valence band's photoexcited holes also have a much weaker effect on the Ti $L_{2,3}$ edge since the probed Ti $2p$ states predominantly compose the conduction band (**Figure 3a**). The holes therefore only change the screening and angular momentum components of the core-valence excitons in the X-ray spectrum without adding or blocking new transitions. The theoretical differential absorption in **Figure 7a**, as compared to thermal and hot electron changes, does demonstrate that adding holes to the calculation should have a measurable effect. Namely, introducing holes to a photoexcited electrons-only model produces an increase in the differential absorption largely at 462 eV and across the spectrum.

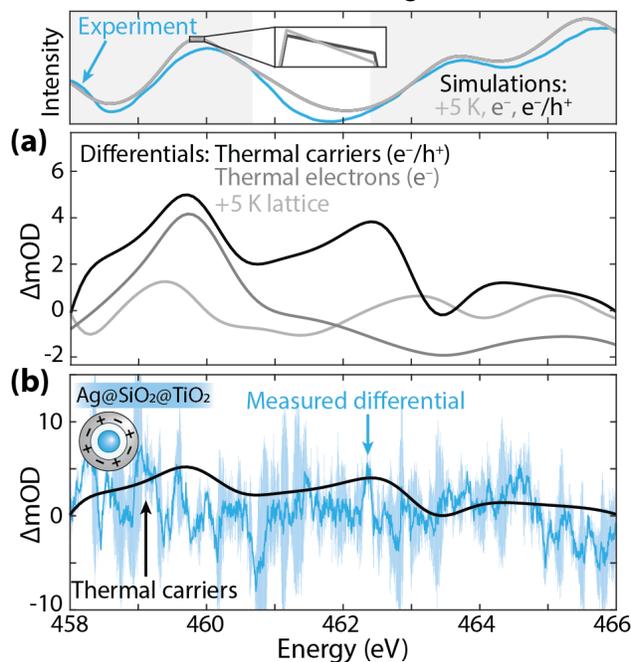

**Figure 7. Thermalized electron plus hole effects on the Ti L$_{2,3}$ edge in Ag@SiO$_2$@TiO$_2$.** (Top) Raw spectra for reference. The spectral range analyzed by the MSE is shaded. (a) Simulated differential spectra for a +5 K lattice expansion (light grey), thermalized electrons (grey), and thermalized carriers (black). The thermalized carriers are simulated at each respective band edge. (b) Thermalized carriers compared to the Ag@SiO$_2$@TiO$_2$ measured differential spectrum. The blue shaded region depicts the experiment's standard deviation.

The Ag@SiO$_2$@TiO$_2$ nanoparticles' measured differential absorption is compared to the simulated differential for thermalized carriers in **Figure 7b**. Calculating the differential absorption spectra and corresponding MSE for thermalized electrons, thermalized electron-hole pairs, and photothermal effects reveals that there is no statistical difference in the MSE for the three calculations, despite the changes in the spectra between these three photoexcited effects (**Figure S2c**). The pre-edge change is well-captured in **Figure 7b**, but this alone is not enough to signify a statistical difference from the other models. The photo-charging indicates that the plasmonic resonance energy transfer effect is present as in previous reports, but a less noisy spectrum or a higher excitation density, which would exaggerate the pre-edge and hole effects, would be needed to differentiate all three models. However, **Figure 7a** indicates that the separation of electrons, holes, or electrons plus holes from photothermal effects should be possible in a quasi-equilibrium, photomodulated experiment.

**CONCLUSION**

We used a set of plasmonic core-shell nanoparticles to test if a BSE-based analysis can differentiate heating, hot electron, and thermalized carrier effects in quasi-equilibrium, photomodulated X-ray absorption experiments. The lattice temperature and hot carrier energy were successfully separated and analyzed within a noisy experimental spectrum. Separating a thermalized electron and hole carrier distribution was not as successful, although this outcome is mainly due to a lower experimental signal and a spectral cancellation unique to the Ti $L_{2,3}$ edge. The BSE method proposed appears accurate enough to allow non-time-resolved X-ray beamlines to determine electron and hole effects, greatly expanding the realm of photoexcited studies. This is a particularly important advance for systems where defects, hot carrier effects, and junctions that control transport and surface catalysis through steady-state distributions are difficult to study with ultrafast spectroscopy. Based on this analysis, future photomodulated spectroscopy measurements should aim for a photomodulated or differential signal-to-noise ratio >2 or an average differential signal of ~5 mOD. The signal-to-noise ratio in this work is 0.5, determined by dividing the average differential signal (1.4 mOD in **Figure 6c**) by the noise root-mean-squared (2.8 mOD in **Figure 6d**). Ultrafast studies have larger signal-to-noise ratios due to a higher impulse response, carrier density, and spectral chopping rate frequency, apparent in **Figure S1**. However, measuring quasi-equilibrium distributions requires a careful balance of the excitation rate, recombination rate, and photomodulation time. The results of our paper therefore give technical guidelines for measuring simultaneous electron, hole, and thermal quasi-equilibrium populations.

**METHODS**

**Core-Shell Nanoparticle Synthesis and Characterization.**

Core-shell metal@(SiO$_2$)@TiO$_2$ nanoparticles were synthesized and characterized previously, and this work is only a theoretical analysis of X-ray spectra for these same particles.[28] All experimental X-ray and UV–Vis spectra were collected at the time of these referenced, initial publications. Aqueous-phase UV–Vis absorption (Shimadzu 2550) measured the localized surface plasmon resonance and TiO$_2$ absorption onset for each particle (**Figure S3**).

**Photodiode Heterojunction Modeling.**

A drift-diffusion model is used to simulate the metal-semiconductor junction for each core-shell nanoparticle design through the Automat FOR Simulation of HETerostructures (AFORS-HET) software v.2.5. This numerical simulation software uses a 1D drift-diffusion model based on self-consistent solutions to the Poisson equation to model the band bending, carrier tunneling, and junction properties.[54] See the Supporting Information for input parameters and the built-in field calculation.

**X-ray Absorption Spectroscopy.**

The National Synchrotron Radiation Research Center (NSRRC) in Hsinchu, Taiwan, collected all Ti L edge X-ray spectra at the BL20A1 beamline in total electron yield mode (reflection geometry) depicted in **Figure S4**. Each specimen was mounted on conductive Cu tape without surface treatment. All secondary electrons were collected to generate the detected signal. The total Xe lamp power density was ~200 mW/cm$^2$ at the sample and ~10 mW/cm$^2$ across each plasmon resonance energy range. The photon flux was measured 1 m away from the lamp with an initial power of 500 W. The lamp was spectrally filtered to irradiate the sample with >400 nm light or below the TiO$_2$ bandgap, and non-AR coated optics were used as the entrance windows. The X-ray spectra were collected with a (lamp off), (lamp on), (lamp off), (lamp on) sequence for 48 s acquisition per spectrum. The X-ray analysis software at the beamline was used to subtract the background (X-ray scattering and electron emission) using a straight baseline fit below the absorption rising edge.

*Ab Initio* **X-ray Theory.**

The X-ray absorption simulation package, Obtaining Core Excitations from the *Ab initio* electronic structure and the NIST BSE solver (OCEAN), was used to model the plasmonic and excited-state properties in TiO$_2$.[55,56] The package's workflow has been described previously.[36,37] A variable-cell relaxation of the TiO$_2$ anatase crystal structure was initially performed to define the suitable cell parameters. As part of the workflow, Quantum ESPRESSO[57,58] performed density functional theory (DFT) to calculate the ground-state electronic structure using a plane-wave basis set and a 350 Ry cutoff energy. The DFT used Trouiller-Martins norm-conserving pseudopotentials, calculated using a Perdew-Wang local density approximation (LDA). A 16x16x12 *k*-point mesh was used with 248 total bands. The macroscopic dielectric constant was set to 5.62 for TiO$_2$ anatase.[59] The Haydock solver is used to calculate all X-ray spectra. The spin-orbit coupling for all BSE calculations was fixed as 4.5 eV. See the Supporting Information for input parameters of the cutoff energy convergence, BSE screening, variable-cell relaxation, and band structure calculations.

The standard OCEAN code can interpret static lattice heating by re-running the DFT and BSE calculations for lattice parameters that simulate an isotropic thermal lattice expansion. However, our previous reports discuss the modified BSE code that simulates excited-state electrons and holes.[26,36,37] The standard code is modified to output the band structure as a usable *k*-point mesh array with defined energy values for each *k*-point. This array is then evaluated and modified to include an excited-state carrier population through state filling. Specifically, photoexcited electrons are simulated by blocking available transitions in the conduction band while holes are simulated by opening or making states available in the valence band. It is worth noting that the state filling fully occupies each state whereas a partial occupation would more accurately depict a carrier density. We use an iterative approach to state filling in MATLAB, which fills all conduction states up to a specified energy or opens valence states for holes. However, this method is complicated for band structures with degenerate valleys across *k*-space because the *k*-point mesh in OCEAN is unsorted. In other words, degenerate valleys would be unavoidably filled with excited-state electrons, and the simulated excitation would not be momentum-specific.

After the X-ray absorption spectra are calculated, they are broadened to account for the experimental lifetime broadening of the $TiO_2$ Ti $L_{2,3}$ edge.[35] The theoretical differential absorption is then calculated as the log of the excited-state spectrum divided by the ground-state spectrum. The mean squared error (MSE) between the calculation and experiment is calculated using MATLAB's 'goodnessOfFit' function. See the Supporting Information for the lattice expansion parameters, state-filling simulations/scripts, spectral broadening procedure, and MSE calculations.

## AUTHOR INFORMATION

**Corresponding Author**

*Scott K. Cushing. Email: scushing@caltech.edu

**Present Addresses**

[1]Division of Chemistry and Chemical Engineering, California Institute of Technology, Pasadena, CA 91125, USA.

[2]Department of Physics, Tamkang University, New Taipei City 251301, Taiwan

[3]Department of Chemistry, National Taiwan University and Advanced Research Center for Green Materials Science and Technology, Taipei 10617, Taiwan

[4]Department of Chemical Engineering, University of Massachusetts Amherst, Amherst, MA 01003–9303, United States

**Author Contributions**

L.D.P. and S.K.C. conceived the research outlook and direction. L.D.P. performed the data analysis and theoretical modeling/simulations. L.D.P. interpreted the results and wrote the manuscript. W.L. supported the theoretical development, and data interpretation, and writing. S.K.C. synthesized and characterized the core-shell nanoparticles. C.L.D. and R.S.L. collected the experimental X-ray data. N.W. directed initial project direction and collaboration. S.K.C. acquired project funding and directed the project. All authors have given approval to the final version of the manuscript.

**Funding Sources**

A portion of this work was supported by the Liquid Sunlight Alliance, which is supported by the U.S. Department of Energy, Office of Science, Office of Basic Energy Sciences, Fuels from Sunlight Hub under Award Number DE-SC0021266. L.D.P. is supported by an NSF Graduate Research Fellowship under Grant Number DGE-1745301. W.L. acknowledges further support from the Korea Foundation for Advanced Studies. The Resnick Sustainability Institute at the California Institute of Technology supports the Resnick High Performance Computing Center. R.S.L. acknowledges financial support by the National Science and Technology Council of Taiwan (NSTC 109-2113-M-002-020-MY3), and the " Advanced Research Center

For Green Materials Science and Technology" from The Featured Area Research Center Program within the framework of the Higher Education Sprout Project by the Ministry of Education (112L9006).

**ACKNOWLEDGEMENTS**

We thank Jonathan Michelsen and Hanzhe Liu for providing guidance and MATLAB scripts to perform the OCEAN calculations. Additionally, we thank Alex Krotz for modifying the OCEAN code to incorporate photoexcited holes. The computations presented here were conducted in the Resnick High Performance Computing Center, a Resnick Sustainability Institute facility at the California Institute of Technology. This research also used resources of the National Energy Research Scientific Computing Center, a DOE Office of Science User Facility supported by the Office of Science of the U.S. Department of Energy under Contract No. DE-AC02-05CH11231 using NERSC award BES-ERCAP0024109.

**Notes**
The authors declare no competing financial interest.

# Supporting Information:

# Determining Quasi-Equilibrium Electron and Hole Distributions of Plasmonic Photocatalysts using Photomodulated X-ray Absorption Spectroscopy

Levi D. Palmer[1], Wonseok Lee[1], Chung Li Dong[2], Ru-Shi Liu[3], Nianqiang Wu[4], Scott K. Cushing[1],*

[1]Division of Chemistry and Chemical Engineering, California Institute of Technology, Pasadena, CA 91125, USA.
[2]Department of Physics, Tamkang University, New Taipei City 251301, Taiwan
[3]Department of Chemistry, National Taiwan University and Advanced Research Center for Green Materials Science and Technology, Taipei 10617, Taiwan
[4]Department of Chemical Engineering, University of Massachusetts Amherst, Amherst, MA 01003–9303, United States
*Corresponding author. Email: scushing@caltech.edu

**Table of Contents**

## 1. Interpreting Previous Ultrafast X-ray Spectra of Anatase TiO$_2$.

The excited-state Bethe-Salpeter equation (BSE) approach is tested on previous ultrafast X-ray measurements of anatase TiO$_2$ collected in ref. 1 (**Figure S1**).[1] This test notably benchmarks the theory's accuracy in modeling carrier distributions and potential discrepancies introduced using the TiO$_2$ anatase phase approximation. **Figure S1a** compares the Ti L$_{2,3}$ transient spectrum 15 ps before and 1 ps after photoexcitation to the ground-state simulation performed in this work. Femtosecond carrier dynamics in TiO$_2$ are simpler to model than plasmonic quasi-equilibrium hot carrier distributions because the transient carrier populations relax as a function of time after direct photoexcitation. Direct photoexcitation also avoids complex photothermal and carrier trapping effects occurring in the steady state. **Figure S1b** compares the simulation and experiment for ultrafast X-ray spectra of anatase TiO$_2$. Fully thermalized electrons and holes are simulated with equal contribution. The spectral features are notably similar but the effects at the L$_3$ t$_{2g}$ and e$_g$ peaks are under- and overapproximated, respectively. This result indicates a relatively strong agreement between the experiment and theory in the advent of well-defined ultrafast carrier dynamics.

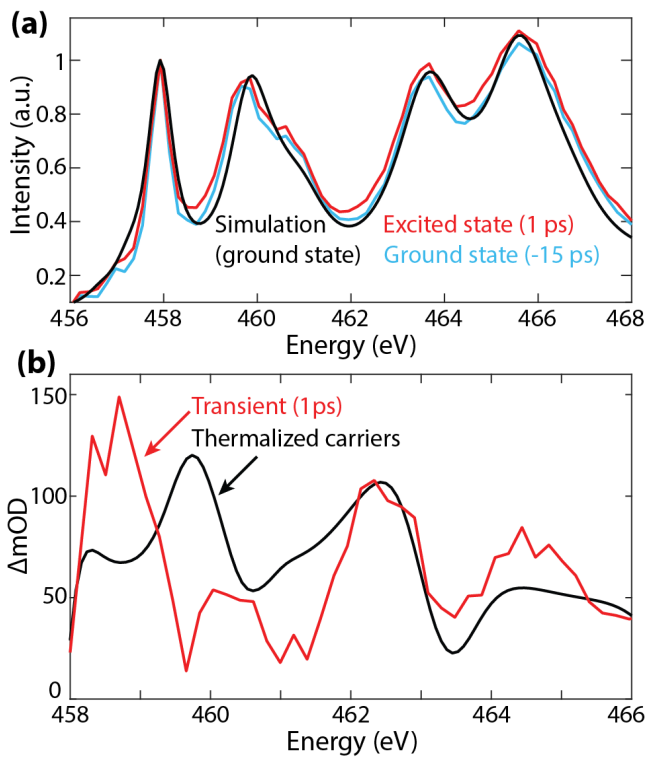

**Figure S1. Simulating carriers in ultrafast Ti L$_{2,3}$ edge X-ray spectra of anatase TiO$_2$.** (a) Ground- and excited-state (-15 ps, blue, and 1 ps, red, before/after photoexcitation) experimental spectra of anatase TiO$_2$ compared to the simulation in this work (black). (b) The 1 ps transient differential of anatase TiO$_2$ modeled with thermalized electrons and holes using the *ab initio* approach. All experimental spectra were measured in ref. [1].

## 2. Quantifying the Fit between Experiment and Theory using Mean Squared Error (MSE)

To quantify the agreement between the simulated and experimental spectra, the mean squared error (MSE) is calculated for each simulated spectrum. The MSE, $\frac{\Sigma_x(y_{simulated} - y_{measured})^2}{N}$, is given by the difference in the spectral intensity for the simulated ($y_{simulated}$) and measured ($y_{measured}$) spectra at each point in energy normalized to the total number of analyzed points in the spectrum (N). The MSE represents the average squared difference between the experiment and theory across the spectral range.

For Au@SiO$_2$@TiO$_2$, the differential X-ray spectrum for each simulated lattice temperature (ΔT of 0, 2.5, 5.0, 10, 15, and 20 K) was calculated, as shown in **Figure S2a**. The MSE for each temperature was fit using a quadratic regression model ($R^2$ = 0.99) to determine the lattice temperature of the Au@SiO$_2$@TiO$_2$ nanoparticles. Because the quadratic regression fits the quadratic relationship of the MSE (mOD$^2$) as a function of temperature (K), the vertex reflects the best match between the experiment and theory, giving an estimate of the photothermal temperature. The fit minimum at 2.5 K has a MSE of ~5 mOD$^2$, or the average difference between the simulated and measured spectrum is ~2 mOD across the spectral range, rounding to the first significant digit.

For Au@TiO$_2$, the MSE prediction and quadratic regression model ($R^2$ = 0.95) were again used to determine the most likely combination of heating and hot electron occupation. The ~5 mOD$^2$ MSE minimum in the quadratic regression suggests that hot electrons exist up to 0.3 eV above the CBM when including a 14 K lattice expansion, where both temperature and electron energy are optimized by minimizing the total MSE (**Figure S2b**). This MSE again equates to an average difference of ~2 mOD between the simulated and measured spectra.

For Ag@SiO$_2$@TiO$_2$, the MSEs for a +5 K lattice expansion, thermalized electrons, and thermalized carriers are displayed. There was no determined significant difference between the three simulations although thermalized electron-hole pair reportedly exist in the TiO$_2$ layer.

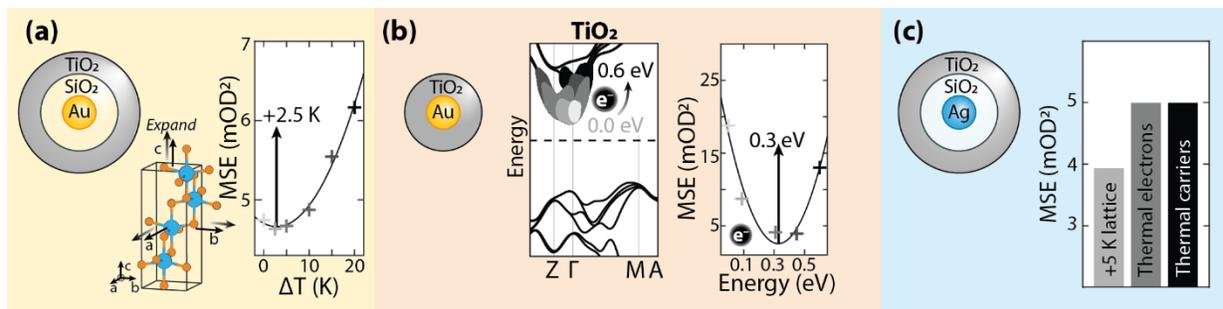

**Figure S2. Mean square error (MSE) analysis for the metal@(SiO$_2$)@TiO$_2$ nanoparticle X-ray simulations.** The MSE between the simulated and measured spectra in Figures (a) 5, (b) 6, and (c) 7 within the main text. The MSEs in (a) and (b) are fit to a quadratic regression model with the fit's vertex at 2.5 K and 0.3 eV, respectively. The MSE analysis in (b) contains both the hot electron simulation and a 14 K lattice expansion as shown by the spectra in **Figure S11**.

## 3. Core-Shell Nanoparticle UV–Visible Spectra.

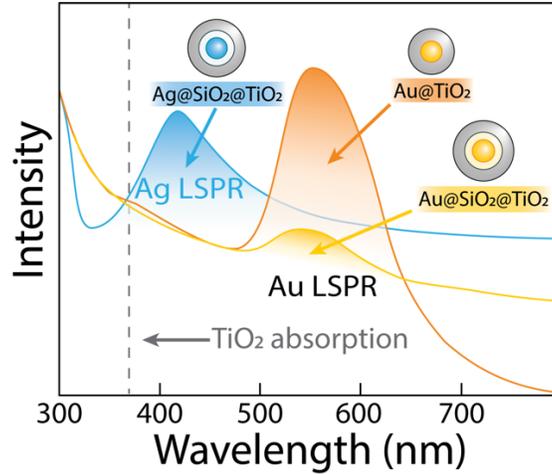

**Figure S3. UV–visible absorption spectra for the metal@(SiO$_2$)@TiO$_2$ nanoparticles.** The intensity is scaled to the TiO$_2$ absorption onset to depict the relative position between each localized surface plasmon resonance (LSPR) and the TiO$_2$ absorption.

## 4. Photodiode Heterojunction Input Parameters.

| Table S1. Input Heterojunction Materials Properties | | | | | |
|---|---|---|---|---|---|
| Parameter | Units | Ag | Au | SiO$_2$ | TiO$_2$ |
| Thickness | [nm] | 15 | 15 | 10 | 20 |
| Dielectric constant | [–] | 1200 | 1200 | 11.9 | 11.9 |
| Electron affinity | [eV] | – | – | 1.0-2.0* | 4.3 |
| Fermi Level | [eV] | -5.3 | -5.2 | – | -4.68 (calc.) |
| Band gap | [eV] | 0.001 | 0.001 | 9.1 | 3.2 |
| Cond./val. band density | [cm$^{-3}$] | 1E22/1E22 | 1E22/1E22 | 2.9E19/2.7E19 | 1E21/1E21 |
| Electron/hole mobility | [cm$^2$/Vs] | 1107/424.6 | 1107/424.6 | 0.01/0.001 | 0.1/0.001 |
| Acceptor/donor concentration | [cm$^{-3}$] | – | – | 0/100* | 0/5E14 |
| Electron/hole thermal velocity | [cm/s] | 1E7/1E7 | 1E7/1E7 | 1E7/1E7 | 1E7/1E7 |
| Layer density | [g*cm$^{-3}$] | 10.5 | 19.3 | 2.7 | 3.84 |

*The electron affinity and dopant concentration of SiO$_2$ should be 0.8 eV and 0 cm$^{-3}$, respectively. However, different values were used to avoid calculation instabilities from the conduction band approaching the vacuum level.

| Table S2. Output Heterojunction Results | | | | |
|---|---|---|---|---|
| Result | Units | Au@SiO$_2$@TiO$_2$ | Au@TiO$_2$ | Ag@SiO$_2$@TiO$_2$ |
| Schottky barrier | [eV] | 0.73 (SiO$_2$@TiO$_2$ interface) | 0.94 | 0.79 (SiO$_2$@TiO$_2$ interface) |
| Junction Fermi level | [eV] | -5.3 | -5.2 | -5.3 |
| Built-in field (in dielectric) | [V/cm] | 1.7E+05 | 2.6E+05 | 2.1E+05 |

We calculate the average built-in electric field ($F$) by calculating the difference in the metal and TiO$_2$ Fermi levels before forming a junction and dividing this difference by the total dielectric and semiconductor thickness ($t$) for TiO$_2$ and SiO$_2$ following equation S1:

$$F = (E_{F,TiO_2} - E_{F,metal})/t \qquad (S1)$$

## 5. Energy-Dependent Broadening of Simulated Spectra.

The OCEAN code uniformly broadens all simulated X-ray spectra using a set energy broadening input. The calculations in this work include 0.1 eV broadening for each output X-ray absorption spectrum. The calculated spectrum from OCEAN is then manually broadened in MATLAB to account for the inherent lifetime broadening of the core-level X-ray transition. This lifetime, or energy-dependent, broadening is a result of the energy-dependent loss function and the inelastic mean free path of the core electron and core hole in TiO$_2$. The spectra are manually broadened by convoluting them with Lorentzian functions with 1.05 (L$_3$ t$_{2g}$), 1.55 (L$_3$ e$_g$), 1.85 (L$_2$ t$_{2g}$), and 2.25 (L$_2$ e$_g$) eV bandwidths. The average broadening of 2.05 and 1.3 eV for the L$_2$ and L$_3$ edges has a ratio of 3:1.9, comparable to the reported 3:2 lifetime broadening ratio for TiO$_2$ anatase.

To accurately model the intensity of the differential spectral features, the intensity of the calculated spectra was also manually modified in MATLAB. The intensity of the convoluted Lorentzian function during the spectral broadening was manually adjusted for each major peak in the experiment to match the ground-state intensity. These peaks include the t$_{2g}$ and e$_g$ peaks for both the L$_3$ and L$_2$ edges. The relative intensity multiplier used for each peak was 1 (L$_3$ t$_{2g}$), 1.65 (L$_3$ e$_g$), 1.35 (L$_2$ t$_{2g}$), and 2.76 (L$_2$ e$_g$) a.u.

The post-edge region was also manually broadened by 5.0 eV and amplified by 3 a.u. However, the differential spectrum in this region was not interpreted due to the sample charging in the experimental measurements. This large (and somewhat arbitrary) manual modification of the post-edge region is a testament to the fact that OCEAN models core-level features on/before the edge better than post-edge features and extended X-ray absorption fine-structure.

## 6. Raw, Experimental X-ray Spectra and Charging Effects.

All experimental data was collected at the BL20A1 beamline (National Synchrotron Radiation Research Center in Hsinchu, Taiwan) in total electron yield mode (reflection geometry) depicted in **Figure S4**.

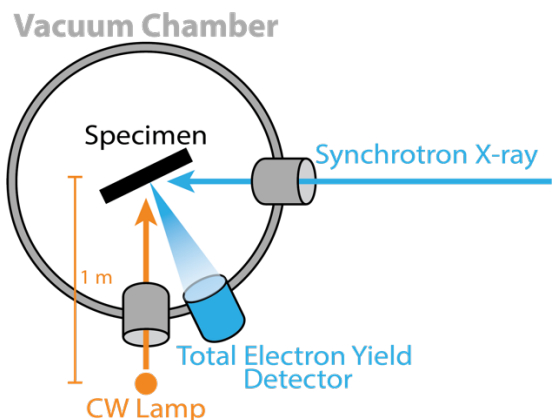

**Figure S4. Photomodulated X-ray absorption spectroscopy in total electron yield detection geometry.** A continuous-wave (CW) lamp photoexcites the specimen from 1 meter away. The incident X-rays probe the dynamics with the lamp on/off, and a total electron yield detector collects electron scattering to emulate X-ray absorption spectroscopy.

The raw, experimental data show spectral intensity fluctuations due to charging for both Au@TiO$_2$ and Ag@SiO$_2$@TiO$_2$ (**Figure 4a**). The influence of charging is apparent by the greater intensity of the light on spectrum (orange or blue) compared to the light off spectrum (black). The Au@SiO$_2$@TiO$_2$ nanoparticles do not reflect charging signatures, which validates the control by suggesting the SiO$_2$ layer effectively blocks hot electron transfer from Au to TiO$_2$. Sample charging is measurable in the total electron yield geometry because the surface charging causes the detected electrons to be acquired more or less efficiently, depending on the charge type, due to Coulombic repulsion.

**Normalization:** The spectra in **Figure 4a** were normalized to the X-ray edge onset maximum near 458 eV to correct for the spectral charging artifacts (**Figure 4b**). This is indicated by a grey dashed line to guide the eye. Using the normalized spectra, the differential absorption was calculated by $\Delta \text{mOD} = \log_{10}\left(\frac{\text{light on}}{\text{light off}}\right) * 10^3$ for each of the two data sets (**Figure 4c**) and averaged (**Figure S5**). The spectra were only interpreted between 458 and 466 eV to avoid regions that may have been affected by the charging normalization. These spectra are overlaid to have a better direct comparison of the differential features and intensities (**Figure S5**).

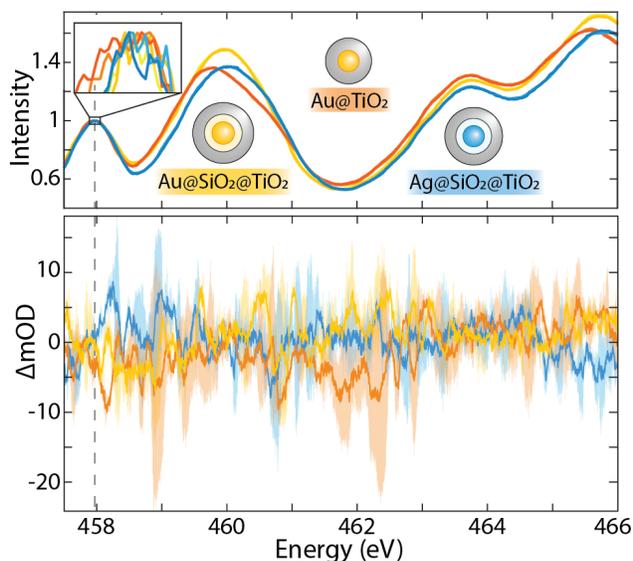

**Figure S5. Photomodulated X-ray spectra of the Ti L$_{2,3}$ edge.** (Top) Experimental ground-state and photomodulated Ti L$_{2,3}$ edge spectra and (Bottom) differential spectra for each core-shell nanoparticle system: Au@SiO$_2$@TiO$_2$ (yellow), Au@TiO$_2$ (orange), and Ag@SiO$_2$@TiO$_2$ (blue). The ground-state spectra are depicted in a lighter shade of each color, but are only distinguishable in the inset. The differential solid lines depict the average raw differential spectra, and the shading depicts the standard deviation of each data point across two averaged spectra. The grey dashed line denotes the point used for charging (amplitude) normalization.

## 7. Comparing Carrier Excitation and Relaxation Rates in Amorphous TiO$_2$.

With the experiment's 10 mW/cm$^2$ power density ($2.7*10^{16}$ photons/(s*cm$^2$) at 560 nm) and the reported ~45% injection efficiency of hot electrons from Au into TiO$_2$, roughly $1.2*10^{16}$ hot electrons/(s*cm$^2$) inject into TiO$_2$.[2] This corresponds to ~$6*10^5$ hot electrons/s injected into each particle's TiO$_2$ layer, assuming uniform particle packing. Ultrafast measurements indicate that hot electrons in crystalline TiO$_2$ films fully thermalize within 20 – 50 fs,[3] but other steady-state spectroscopic measurements have reported hot electron trapping in amorphous TiO$_2$ surface states that prevents carrier and phonon scattering and extends the carrier cooling time.[4–6]

## 8. Ground-State Calculations.

### 8A. Ground-state DFT (Quantum ESPRESSO): Cutoff energy convergence, variable-cell relaxation, and band structure calculations.

Quantum ESPRESSO (QE), a DFT package, was used to calculate the ground-state electronic structure inputs for the OCEAN X-ray calculations. A variable-cell crystal structure relaxation was used to define the simulated atomic coordinates of $TiO_2$ anatase, and a convergence calculation in QE was used to define the cutoff energy. QE was also separately used to calculate the projected density of states (PDOS) and band structure of $TiO_2$.

**Variable-Cell Relaxation.** A variable-cell relaxation (vc-relax) calculation was used to determine the anatase atomic coordinates by optimizing the unit cell dimensions. The vc-relax calculation was completed with the cell_dofree = 'ibrav' input to maintain consistency in the lattice structure while relaxing (optimizing) the unit cell axis and angles. See **Appendix A** for input parameters.

**Convergence Calculations.** The cutoff energy used for the X-ray simulations was first checked for total energy convergence using QE self-consistent field (SCF) calculations. The SCF calculation (**Appendix A**) calculates the total energy of the unit cell lattice, which is used to determine the calculation's convergence. The convergence is a representation of the decrease in the total energy and a higher calculation accuracy following the increase of the wavefunction's (pseudopotential's) energy. The purpose is to preserve calculation accuracy while reducing computational expense. As shown in **Figure S6**, the 350 Ry cutoff energy used for all OCEAN calculations was sufficiently converged.

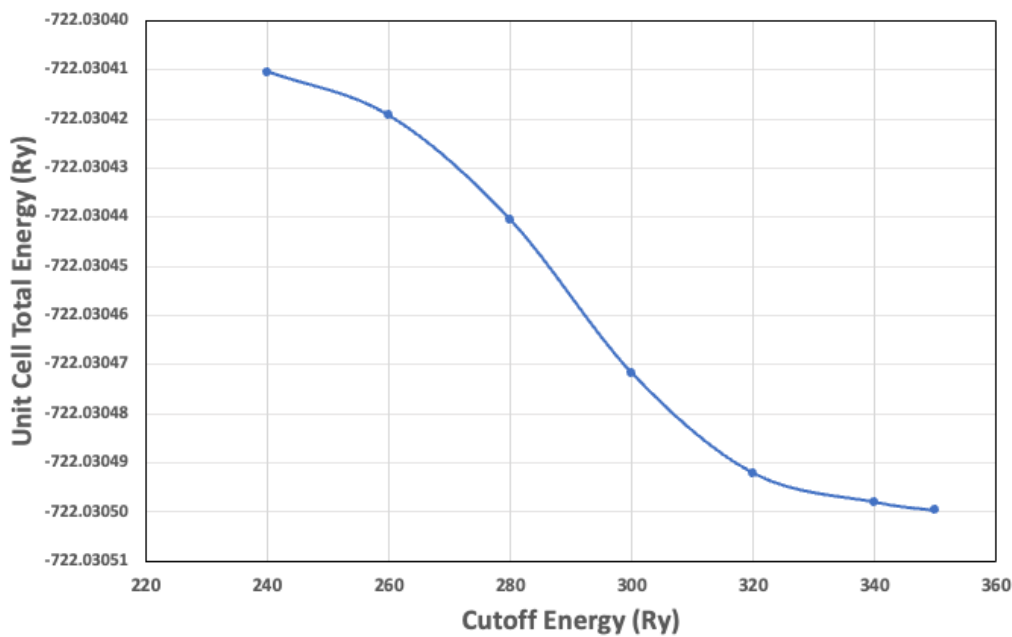

**Figure S6. Cutoff energy convergence.** A $TiO_2$ (anatase) self-consistent field calculation was used to calculate the total unit cell lattice energy at various QE cutoff energy input values.

The *k*-point mesh convergence was similarly confirmed using the 350 Ry cutoff energy. However, the *k*-points largely determine the accuracy of the state-filling calculation, so a large *k*-point mesh are desired regardless of the convergence threshold. A 16x16x12 *k*-point mesh was used for all OCEAN X-ray calculation (**Appendix B**).

**Projected Density of States (PDOS) and Band Structure.** The band structure simulation included the following calculation stages: SCF, non-self-consistent field (NSCF), Bands, and converting/plotting the data. The SCF calculation calculates the wavefunctions for the unit cell used for the density of states calculation, which is extrapolated in *k*-space with the NSCF calculation using a higher *k*-point mesh. The input parameters for each step can be found in **Appendix A**. Plotted in **Figure S7B**, the band structure *k*-path is defined in the Bands calculation, following previous literature.[7] The calculated Fermi level is located at the valence band edge when calculated with DFT.

The PDOS simulation included the following calculation stages: SCF, NSCF, and plotting the projected states for each atomic orbital. The PDOS is plotted in **Figure S7C** for the relevant valence states, Ti 3*d* (blue) and O 2*p* (orange). Background shading in **Figure S7C,D** depicts the crystal field-split $t_{2g}$ (blue shading) and $e_g$ (grey shading) states.

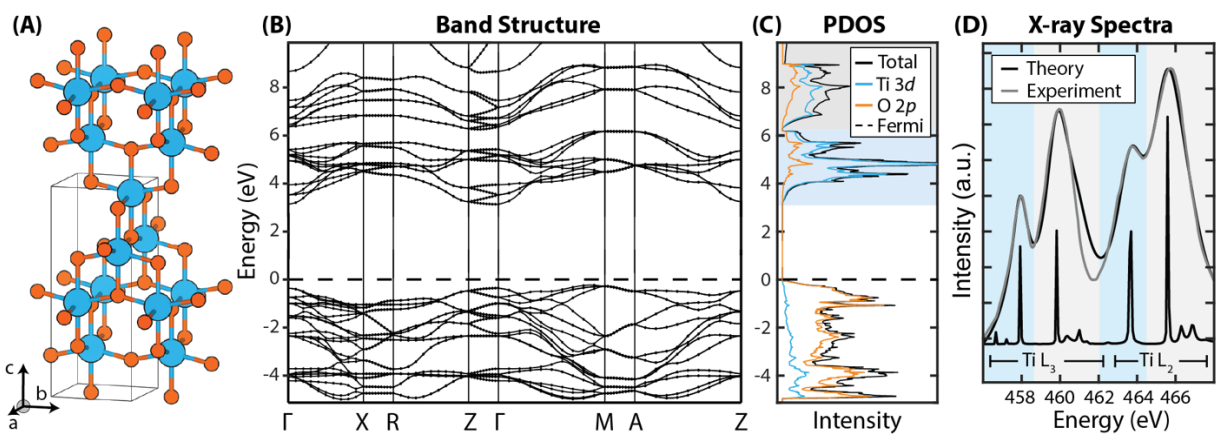

**Figure S7.** *Ab initio* **ground-state calculations of TiO$_2$ anatase.** (A) The variable-cell relaxed crystalline TiO$_2$ anatase structure unit cell with calculations of the (B) band structure, (C) projected density of states (PDOS), and (D) X-ray spectra. A 1 eV scissor shift is applied to extend the bandgap to the experimental value. The band structure and PDOS were calculated using Quantum ESPRESSO, and the calculated X-ray spectra were calculated with OCEAN and are also shown in the main text.

### 8B. Ground-state X-ray theory (OCEAN): BSE, screening, and scissor shift.

To simulate the TiO$_2$ (anatase) L$_{2,3}$ edge, the 'Obtaining Core Excitations from *Ab initio* electronic structure and the NIST BSE solver (OCEAN)' code was implemented. OCEAN is a DFT and GW/BSE approach to simulating core-level electron excitations. The DFT framework is the Quantum ESPRESSO package,[8,9] specified in the OCEAN input file (see **Appendix B**). The DFT stage is first calculated in OCEAN to determine the ground-state electronic structure.

Notably, the OCEAN code uses the BSE to simulate the Coulombic effects of the core-to-valence transition exciton. OCEAN implements a screening stage to simulate the screening of the core-hole by the valence state electrons. The combined BSE and screening stages alongside the angular momentum matrix elements of the X-ray transition Hamiltonian calculate the core-level transition (or X-ray absorption) spectrum.

An additional input, core_offset = .true., was included in the OCEAN calculations here to calculate the core-level shifts. The core-level shifts are important because they produce the Kohn-Sham potentials at each atomic site, which improves the calculation's accuracy by accounting for the unique core-level shifts at each atom and screening radius even if the atoms are equivalent sites. A 1 eV scissor shift was also used to adjust the simulated band gap to the 3.2 eV experimental value, but this additional step did not appear to significantly affect the output X-ray spectrum.

9. **Excited-State X-ray Theory and Statistical Error Calculations: lattice expansion parameters (heating), state-filling simulations, and mean-squared error (MSE) calculations.**

An adiabatic approximation was taken to simulate the photoexcited and quasi-equilibrium dynamics. In other terms, the excited-state dynamics are longer lived than the initial electron field excitation from the photon field.

**9A. Heating (Thermal Lattice Expansion):**

Photoexcited thermal effects are accounted for using the thermal expansion coefficient of $TiO_2$ anatase.[10] The lattice expansion is anisotropic with the two expansion coefficients being 4.469E-06 K$^{-1}$ (a and b directions) and 8.4283E-06 K$^{-1}$ (c direction). Equation S2 describes the calculation of an expanded lattice parameter ($d$) at an elevated temperature ($T$) using the lattice expansion coefficient ($\alpha$) with an assumed ground-state/room temperature at 300 K. The initial lattice constants at 300 K ($d_0$) were 7.052 Bohr (a and b directions) and 17.81 Bohr (c direction).

$$d = (\alpha * d_0 * (T - 300\ K)) + d_0 \qquad (S2)$$

The lattice expansion was calculated and simulated for 302.5 K, 305 K, 310 K, 315 K, and 320 K. The percent expansion for each temperature is shown in **Table S3**. The spectra calculated with these OCEAN input lattice constants are found in the main text **Figure 5a**.

| Table S3: TiO$_2$ Anatase Lattice Expansion Calculation Results | | | | | |
|---|---|---|---|---|---|
| Temperature (K) | a (Bohr) | b (Bohr) | c (Bohr) | a % change | c % change |
| 300 | 7.05288 | 7.05289 | 17.81326 | – | – |
| 302.5 | 7.05296 | 7.05297 | 17.81363 | 0.0011 | 0.0021 |
| 305 | 7.05304 | 7.05305 | 17.81401 | 0.0022 | 0.0042 |
| 310 | 7.05320 | 7.05321 | 17.81476 | 0.0044 | 0.0084 |
| 315 | 7.05336 | 7.05337 | 17.81551 | 0.0067 | 0.0126 |
| 320 | 7.05352 | 7.05352 | 17.81626 | 0.0089 | 0.0168 |

## 9B. Hot Electrons Simulated in TiO$_2$ Anatase:

The OCEAN v2.5.2 source code was previously modified to include photoexcited carriers using state-filling simulations.[11] Following these modifications, the OCEAN code automatically outputs an array file after the CNBSE stage that contains unsorted valence states in k-space defined as occupied (contains an electron = 1) or unoccupied (no electrons = 0). These states are the valence band and conduction band, respectively, and the number of states is set using the *k*-point mesh defined in the OCEAN input file (16x16x12 = 3072 states in this work). **Figure S8A,B** depict the array file output from the OCEAN code where the *k*-points are not sorted along a specific *k*-path following high-symmetry *k*-points. Each *k*-point has an associated energy as shown in **Figure S8A**. Then, the *k*-points can be manually sorted into the same high-symmetry path used for the band structure (**Figure S8C**). Sorting the *k*-points into the band structure is an essential step for later visualizing the simulated carrier distributions.

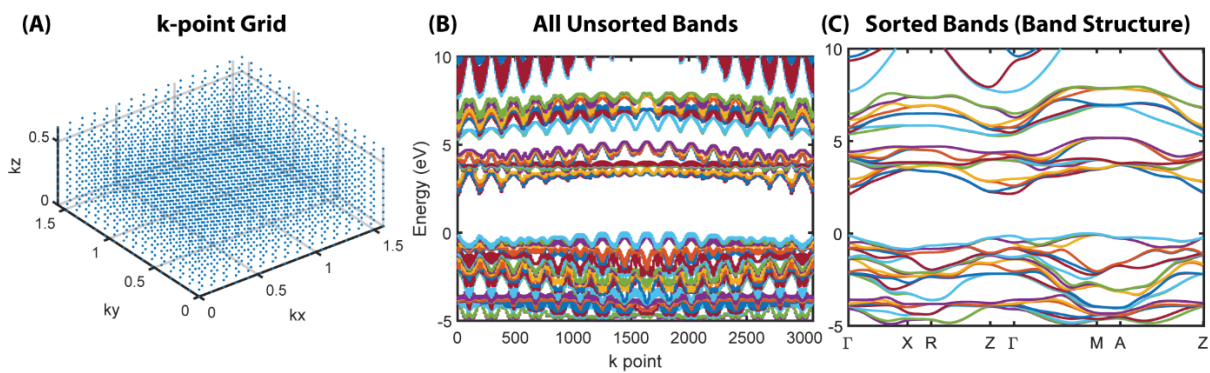

**Figure S8. Unsorted and sorted *k*-points for the array file and TiO$_2$ anatase band structure.** (a) The *k*-point grid depicts the number of *k*-points simulated in reciprocal space. (b) The unsorted *k*-points plotted by energy. The 1 eV scissor shift of the band gap is not included, and the Fermi level (0 eV) is defined at the valence band edge. (C) The sorted *k*-points along the *k*-path of the labelled high-symmetry *k*-points.

The state-filling model takes the valence states defined as '0' or '1' and redefines their values to simulate photoexcited electrons and/or holes. For the hot electron state-filling algorithm in this work, the electronic states up to a specified value are all fully filled. Hot electrons up to 0.0 (fully thermalized), 0.1, 0.3, 0.45, and 0.6 eV above the conduction band minimum were simulated in this work, shown in the main text. The input hot electron state-filling simulations used for all calculations are shown in **Figure S9**. The electron distributions were interpolated onto the band structure because OCEAN uses the unsorted *k*-points as an input (**Figure S8B**). Note that this state-filling method would not typically be *k*-point selective for other band structures; however, all conduction band states at or below 0.6 eV for $TiO_2$ anatase have occupations at or near the gamma point. Therefore, the band structure allows for the simulation of hot electrons at the gamma point alone as compared to other, more complicated band structures that would distribute the hot electrons in *k*-space.

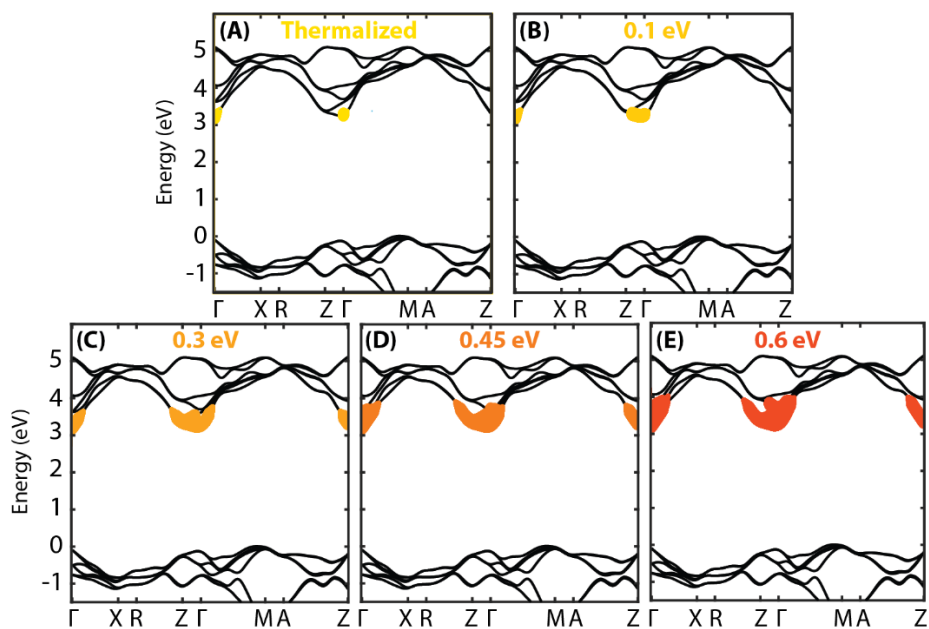

**Figure S9**. **Hot electron state-filling simulations**. Electronic states simulated with hot electrons (colored in yellow to orange gradient) filling from the conduction band minimum up to (A) 0.0 eV thermalized/one conduction state filled, (B) 0.1 eV, (C) 0.3 eV, (D) 0.45 eV, and (E) 0.6 eV. Each state-filling occupation was separately input into OCEAN. A 1 eV scissor shift is included to accurately depict the simulation input. The bands were interpolated in *k*-space to project the simulated state-filling onto the band structure.

## 9C. Electrons and Holes Simulated in $TiO_2$ Anatase:

Similar to the procedure for hot electrons alone, fully thermalized electrons and holes were simulated with OCEAN state-filling at the conduction band minimum and valence band maximum. This approach was taken to simulate the effect of plasmon-induced dipole coupling on the photomodulated X-ray spectra. The energies of the thermalized electrons and holes were at the conduction and valence band edges (only one state filled). **Figure S10** depicts the state-filling simulation with the electrons and holes in each respective band. The M → Γ indirect transition is assumed due to the indirect band gap and lowest transition energy for anatase $TiO_2$.

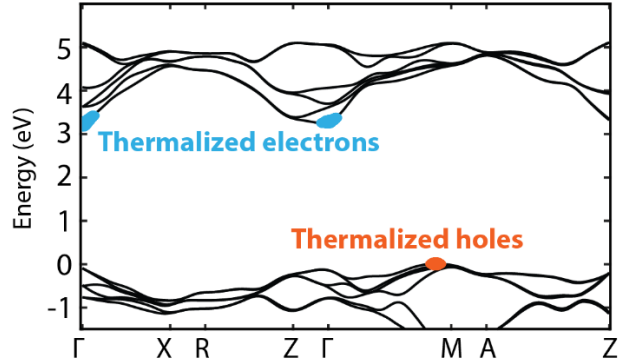

**Figure S10. Fully thermalized electron and hole state-filling simulations.** Thermalized electrons and holes were input into the OCEAN simulations as the state-filled band structure shown. A 1 eV scissor shift is included to accurately depict the simulation input. The bands were interpolated in *k*-space to project the simulated state-filling onto the band structure.

## 9D. Mean Squared Error (MSE) Calculations

The MATLAB mean squared error (MSE) 'goodnessOfFit' function was used to quantify the difference between the simulated differential spectra from OCEAN and the raw, photomodulated experimental data. Specifically, both the simulated and experimental differential spectra were compared in the 458 – 460.6 eV and 462.4 – 466 eV energy ranges of interest using the 'MSE' fit function. These energy windows were selected because of the accuracy of the $TiO_2$ anatase approximation for the experimentally measured amorphous $TiO_2$ (**Figure 3**). For the lattice heating simulations, the intensity of the differential spectrum was unmodified and determined solely by the intensity of the simulated X-ray spectrum output. For the state-filling simulations, the intensity of the spectrum was first chosen to minimize the MSE between the simulated and experimental X-ray differential spectra. This is essential because the OCEAN code cannot accurately predict the exact differential intensity (ΔOD) of carriers as partial state occupations, and a low carrier density exists experimentally. The state-filling simulation's differential intensity was then normalized to the number of simulated carriers to account for intensity fluctuations caused by the increased carrier density. The values selected for normalizing the hot electron state-filling spectral intensity are shown in **Table S4**, and the total intensity of the spectrum per carrier simulated is 4,500,000.

| Table S4: $TiO_2$ Anatase State-Filling Spectrum Normalization | | |
|---|---|---|
| Number of Simulated States Filled | Spectrum Multiplier | Total Intensity per Carrier Simulated |
| 1 (Thermalized at CBM) | 400000 | 400000* |
| 20 (~0.1 eV above CBM) | 200000 | 4000000 |
| 150 (~0.3 eV above CBM) | 26666 | 4000000 |
| 300 (~0.45 eV above CBM) | 13333 | 4000000 |
| 566 (0.65 eV above CBM) | 7067 | 4000000 |

*The intensity for one (1) simulated electron was found to be notably highly non-linear as compared to the intensity for 20 – 566 electrons in the conduction band. The total carrier intensity multiplier was arbitrarily set to 400,000 for this fully thermalized electron simulation.

After the MSE was calculated for each simulation, the lattice heating and hot electron state-filling simulations were further examined using a quadratic regression of the MSEs to define the optimal/minimized lattice temperature and hot electron energy. MATLAB's 'fitlm' function and a 'quadratic' regression model were used to determine the R-squared value of the quadratic fit, and the function's minimum was referenced as the temperature or hot electron energy most representative of the experiment's conditions.

## 9E. Hot Electron X-ray Differential Spectra with Temperature Included (Used for the MSE Analysis)

Following the hot electron X-ray simulations, a 14 K thermal contribution to the spectrum was included. This lattice heating was added for the MSE analysis, which is not reflected in **Figure 6b** in the main text. For clarity and transparency, these simulated differentials are presented in **Figure S11**.

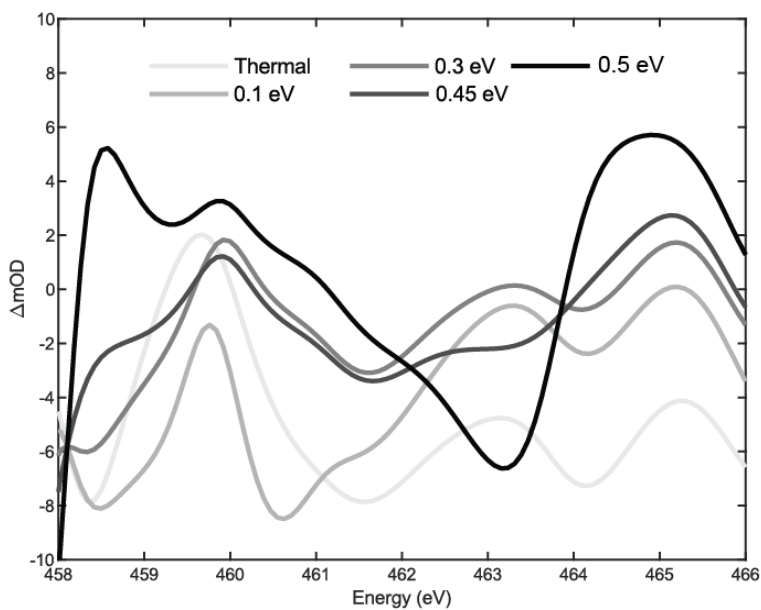

**Figure S11**. **Simulated hot electron differential spectra with temperature included.** The temperature incorporated in the simulation was fixed at 13.5 K with a 5.8 mOD intensity shift (y-axis) and the carrier normalization as highlighted in **Table S4**.

**Appendix A. Example Quantum ESPRESSO Input Files:**

**Self-Consistent Field Input File (example used for band structure calculation):**
```
&control
   calculation = 'scf'
   restart_mode='from_scratch'
   prefix = 'anatase'
   outdir = './outdir'
   pseudo_dir = '-'
/
&system
 ibrav = 0
 nat = 12
 ntyp = 2
 noncolin = .false.
 lspinorb = .false.
 ecutwfc = 220
 occupations = 'fixed'
 smearing = 'gaussian'
 degauss = 0.002
 nspin  = 1
 tot_charge  = 0.0
 nosym = .true.
 noinv = .true.
/
&electrons
 conv_thr = 1.1d-8
 mixing_beta = 0.3
 electron_maxstep = 250
 startingwfc = 'atomic+random'
 diagonalization = 'david'
/
ATOMIC_SPECIES
 Ti  47.86700  ti.fhi.UPF
 O   15.99940  08-o.lda.fhi.UPF
CELL_PARAMETERS cubic
  7.052884613   0.000000000  0.0000000000
  0.000000000   7.052893877  0.0000000000
  0.000000000   0.000000000  17.813258477
ATOMIC_POSITIONS crystal
Ti    0.9989304300    0.9990218340    0.9889160540
Ti    0.4989346150    0.4990157270    0.4889157940
Ti    0.9989331860    0.4990225240    0.2389181510
Ti    0.4989289910    0.9990164210    0.7389132230
O     0.9989302620    0.9990225790    0.1969983610
```

| | | |
|---|---|---|
| O | 0.4989269390 | 0.4990168940 | 0.6969918920 |
| O | 0.9989346180 | 0.4990128120 | 0.4469955950 |
| O | 0.4989312380 | 0.9990212260 | 0.9469954940 |
| O | 0.4989334320 | 0.4990235570 | 0.2808392350 |
| O | 0.9989301330 | 0.9990146590 | 0.7808332590 |
| O | 0.4989345910 | 0.9990164110 | 0.5308359080 |
| O | 0.9989308730 | 0.4990219780 | 0.0308359370 |

K_POINTS (automatic)
 6 6 6 0 0 0

**Non-Self-Consistent Field Input File (example used for band structure calculation):**
&control
   calculation = 'nscf'
   restart_mode='from_scratch'
   prefix = 'anatase'
   outdir = './outdir'
   pseudo_dir = '-'
/
&system
 ibrav = 0
 nat = 12
 ntyp = 2
 noncolin = .false.
 lspinorb = .false.
 ecutwfc = 220
 occupations = 'fixed'
 smearing = 'gaussian'
 degauss = 0.002
 nspin = 1
 tot_charge = 0.0
 nosym = .true.
 noinv = .true.
   nbnd=200
   occupations = 'tetrahedra'
/
&electrons
   conv_thr=1.1d-8
/
ATOMIC_SPECIES
 Ti  47.86700  ti.fhi.UPF
 O   15.99940  08-o.lda.fhi.UPF
CELL_PARAMETERS cubic
  7.052884613   0.000000000  0.0000000000
  0.000000000   7.052893877  0.0000000000
  0.000000000   0.000000000  17.813258477

ATOMIC_POSITIONS crystal
Ti   0.9989304300   0.9990218340   0.9889160540
Ti   0.4989346150   0.4990157270   0.4889157940
Ti   0.9989331860   0.4990225240   0.2389181510
Ti   0.4989289910   0.9990164210   0.7389132230
O    0.9989302620   0.9990225790   0.1969983610
O    0.4989269390   0.4990168940   0.6969918920
O    0.9989346180   0.4990128120   0.4469955950
O    0.4989312380   0.9990212260   0.9469954940
O    0.4989334320   0.4990235570   0.2808392350
O    0.9989301330   0.9990146590   0.7808332590
O    0.4989345910   0.9990164110   0.5308359080
O    0.9989308730   0.4990219780   0.0308359370

K_POINTS (automatic)
 8 8 8 0 0 0

**Variable-Cell Relaxation Input File:**
&control
   calculation = 'vc-relax'
   prefix = 'anatase'
   outdir = './outdir'
   pseudo_dir = './'
   wfcdir = 'undefined'
   tstress = .false.
   tprnfor = .false.
   wf_collect = .true.
/
&SYSTEM
 ibrav = 0
 nat = 12
 ntyp = 2
 noncolin = .false.
 lspinorb = .false.
 ecutwfc = 220
 occupations = 'fixed'
 smearing = 'gaussian'
 degauss = 0.002
 nspin = 1
 tot_charge = 0.0
 nosym = .true.
 noinv = .true.
 nbnd = 248
/
&electrons
   conv_thr=1.1d-8

```
   mixing_beta = 0.3
   electron_maxstep = 250
   startingwfc = 'atomic+random'
   diagonalization = 'david'
/
&ions
/
&cell
  cell_dofree = 'ibrav'
/
ATOMIC_SPECIES
 Ti   47.8670   ti.fhi.UPF
 O    15.9994   08-o.lda.fhi.UPF
CELL_PARAMETERS cubic
  7.05288461300000       0.000000000000000E+000  0.000000000000000E+000
  0.000000000000000E+000  7.05289387700000       0.000000000000000E+000
  0.000000000000000E+000  0.000000000000000E+000  17.8132584770000
ATOMIC_POSITIONS crystal
Ti    0.9989304300    0.9990218340    0.9889160540 0 0 0
Ti    0.4989346150    0.4990157270    0.4889157940 0 0 0
Ti    0.9989331860    0.4990225240    0.2389181510 0 0 0
Ti    0.4989289910    0.9990164210    0.7389132230 0 0 0
O     0.9989302620    0.9990225790    0.1969983610 0 0 0
O     0.4989269390    0.4990168940    0.6969918920 0 0 0
O     0.9989346180    0.4990128120    0.4469955950 0 0 0
O     0.4989312380    0.9990212260    0.9469954940 0 0 0
O     0.4989334320    0.4990235570    0.2808392350 0 0 0
O     0.9989301330    0.9990146590    0.7808332590 0 0 0
O     0.4989345910    0.9990164110    0.5308359080 0 0 0
O     0.9989308730    0.4990219780    0.0308359370 0 0 0
K_POINTS automatic
8 8 8 0 0 0
```

**Projected Density of States File (Run After SCF and NSCF):**
```
&PROJWFC
  prefix= 'anatase',
  outdir= './outdir',
  filpdos= ' anatase_pdos.dat'
/
```

**Band Structure Files:**
**--- Bands Calculation ---**
```
&control
   calculation = 'bands'
   restart_mode='from_scratch'
```

```
   prefix = 'anatase'
   outdir = './outdir'
   pseudo_dir = '-'
/
&system
 ibrav = 0
 nat = 12
 ntyp = 2
 noncolin = .false.
 lspinorb = .false.
 ecutwfc = 220
 occupations = 'fixed'
 smearing = 'gaussian'
 degauss = 0.002
 nspin  = 1
 tot_charge  = 0.0
 nosym = .true.
 noinv = .true.
   nbnd=200
/
&electrons
   conv_thr=1.1d-8
/
ATOMIC_SPECIES
 Ti  47.86700  ti.fhi.UPF
 O   15.99940  08-o.lda.fhi.UPF
CELL_PARAMETERS cubic
   7.052884613   0.000000000  0.0000000000
   0.000000000   7.052893877  0.0000000000
   0.000000000   0.000000000  17.813258477
ATOMIC_POSITIONS crystal
Ti    0.9989304300    0.9990218340    0.9889160540
Ti    0.4989346150    0.4990157270    0.4889157940
Ti    0.9989331860    0.4990225240    0.2389181510
Ti    0.4989289910    0.9990164210    0.7389132230
O     0.9989302620    0.9990225790    0.1969983610
O     0.4989269390    0.4990168940    0.6969918920
O     0.9989346180    0.4990128120    0.4469955950
O     0.4989312380    0.9990212260    0.9469954940
O     0.4989334320    0.4990235570    0.2808392350
O     0.9989301330    0.9990146590    0.7808332590
O     0.4989345910    0.9990164110    0.5308359080
O     0.9989308730    0.4990219780    0.0308359370

K_POINTS {crystal_b}
```

8
  0.0 0.0 0.0 10 !Gamma
  0.5 0.0 0.0 10 !X
  0.5 0.0 0.5 10 !R
  0.0 0.0 0.5 10 !Z
  0.0 0.0 0.0 10 !Gamma
  0.5 0.5 0.0 10 !M
  0.5 0.5 0.5 10 !A
  0.0 0.0 0.5 10 !Z

**--- Preparing Output for 'plotband' ---**
&bands
  outdir='./outdir/'
  prefix='anatase'
  filband='anatase.bands.dat'
/

**--- Plotting the Bands, 'plotband' ---**
anatase.bands.dat
2 18
anatase.bands.xmgr
anatase.bands.ps
7.4320
2 7.4320

**Appendix B. Example OCEAN 2.5.2 Input File:**
ppdir '../'
dft{qe}
para_prefix{ mpirun -np 56 }

################################

nkpt { 16 16 12 }
ngkpt { 16 16 12 }

screen.nkpt { 2 2 2 }
screen.nbands 200

nbands 248

mixing { 0.3 }

acell { 7.052884613  7.052893877  17.813258477 }
rprim {
    1.0 0.0 0.0
    0.0 1.0 0.0
    0.0 0.0 1.0}

ntypat 2
znucl { 22 8 }
zsymb { Ti O}

ppdir { '../' }
pp_list{ ti.fhi
       08-o.lda.fhi }

natom 12
typat { 1 1 1 1 2 2 2 2 2 2 2 2 }

xred {
  0.998930430   0.999021834   0.988916054
  0.498934615   0.499015727   0.488915794
  0.998933186   0.499022524   0.238918151
  0.498928991   0.999016421   0.738913223
  0.998930262   0.999022579   0.196998361
  0.498926939   0.499016894   0.696991892
  0.998934618   0.499012812   0.446995595
  0.498931238   0.999021226   0.946995494
  0.498933432   0.499023557   0.280839235
  0.998930133   0.999014659   0.780833259
  0.498934591   0.999016411   0.530835908

```
   0.998930873   0.499021978   0.030835937}
```

ecut 350
toldfe 1.1d-8
tolwfr 1.1d-16

nstep 250

# Static dielectric const https://aip.scitation.org/doi/pdf/10.1063/1.435102
diemac 5.62

CNBSE.xmesh { 8 8 8 }

opf.fill{ 22 ti.fill }
opf.opts{ 22 ti.opts }

# edge information # number of edges to calculate # atom number, n quantum number, l quantum number
edges{ -22 2 1 }

cnbse.broaden{ 0.1 }

screen.shells{ 4.0 }
cnbse.rad{ 4.0 }

spin-orbit 3.828

scfac 0.8
occopt 1
core_offset .true.
bshift{48} #Sets the number of valence bands in the modified code